\documentclass[journal]{IEEEtran}
\IEEEoverridecommandlockouts
\usepackage{cite}
\usepackage{hyperref}
\usepackage{amsmath,amssymb,amsfonts}
\usepackage{algorithmic}
\usepackage{graphicx}
\usepackage{textcomp}
\usepackage{wrapfig}
\usepackage{xcolor}
    
\usepackage[english]{babel}
\usepackage[utf8]{inputenc}
\usepackage[T1]{fontenc}

\usepackage{calc}
\usepackage{ifthen}
\usepackage{tikz}
\newcommand{\slice}[4]{
  \pgfmathparse{0.5*#1+0.5*#2}
  \let\midangle\pgfmathresult

  \draw[thick,fill=black!10] (0,0) -- (#1:1) arc (#1:#2:1) -- cycle;

  \node[label=\midangle:#4] at (\midangle:1) {};

  \pgfmathparse{min((#2-#1-10)/110*(-0.3),0)}
  \let\temp\pgfmathresult
  \pgfmathparse{max(\temp,-0.5) + 0.8}
  \let\innerpos\pgfmathresult
  \node at (\midangle:\innerpos) {#3};
}

\usepackage{afterpage}
\usepackage{amsmath}
\usepackage{amsthm}
\usepackage{amssymb}
\usepackage{csquotes}
\usepackage{enumitem}
\usepackage{graphicx}
\usepackage{lipsum}
\usepackage{booktabs}
\usepackage[utf8]{inputenc}
\usepackage[english]{babel}
\usepackage{multirow}

\newtheorem{thm}{Theorem}[section]
\newtheorem{defn}[thm]{Definition}
\newtheorem*{remark}{Remark}

\usepackage{listings}
\usepackage[ruled,vlined]{algorithm2e}
\makeatletter
\renewcommand{\SetKwInOut}[2]{%
  \sbox\algocf@inoutbox{\KwSty{#2}\algocf@typo:}%
  \expandafter\ifx\csname InOutSizeDefined\endcsname\relax
    \newcommand\InOutSizeDefined{}\setlength{\inoutsize}{\wd\algocf@inoutbox}%
    \sbox\algocf@inoutbox{\parbox[t]{\inoutsize}{\KwSty{#2}\algocf@typo:\hfill}~}\setlength{\inoutindent}{\wd\algocf@inoutbox}%
  \else
    \ifdim\wd\algocf@inoutbox>\inoutsize%
    \setlength{\inoutsize}{\wd\algocf@inoutbox}%
    \sbox\algocf@inoutbox{\parbox[t]{\inoutsize}{\KwSty{#2}\algocf@typo:\hfill}~}\setlength{\inoutindent}{\wd\algocf@inoutbox}%
    \fi%
  \fi
  \algocf@newcommand{#1}[1]{%
    \ifthenelse{\boolean{algocf@inoutnumbered}}{\relax}{\everypar={\relax}}%
    {\let\\\algocf@newinout\hangindent=\inoutindent\hangafter=1\parbox[t]{\inoutsize}{\KwSty{#2}\algocf@typo:\hfill}~##1\par}%
    \algocf@linesnumbered
  }}%
\makeatother

\usepackage{bm}

\usepackage{tipa}
\usepackage{tcolorbox}
\definecolor{mdgrey}{rgb}{0.8, 0.8, 0.8}
\usepackage[framemethod=tikz]{mdframed}
\usepackage{lipsum}
\newtheoremstyle{defi}
  {\topsep}%
  {\topsep}%
  {\normalfont}%
  {}%
  {\bfseries}%
  {:}%
  {.5em}%
  {\thmname{#1}\thmnote{~(#3)}}%
\theoremstyle{defi}
\newmdtheoremenv{definitioni}{Definition}
\newmdtheoremenv[
hidealllines=true,
leftline=true,
innertopmargin=0pt,
innerbottommargin=0pt,
linewidth=4pt,
linecolor=gray!40,
innerrightmargin=0pt,
]{definitionii}{Definition}
\newmdtheoremenv[
roundcorner=5pt,
innertopmargin=0pt,
innerbottommargin=5pt,
linewidth=4pt,
linecolor=gray!40,
]{definitioniii}{Definition}

\author{Aida Manzano Kharman, Christian Jursitzky, Quan Zhou,\\ Pietro Ferraro, Jakub Marecek, Pierre Pinson, Robert Shorten

\thanks{Aida Manzano Kharman, Christian Jursitzky, Quan Zhou, Pietro Ferraro, Pierre Pinson and Robert Shorten are affiliated with the Dyson School of Design Engineering, Imperial College London. They can be contacted, respectively at the following email adresses aida.manzano-kharman17@imperial.ac.uk, christian.jursitzky21@imperial.ac.uk, q.zhou22@imperial.ac.uk, p.ferraro@imperial.ac.uk, p.pinson@imperial.ac.uk, r.shorten@imperial.ac.uk}
\thanks{ Jakub Marecek is affiliated with the department of Computer Science, Czech Technical University in Prague. He can be contacted at the following email address: jakub.marecek@gmail.com}
\thanks{This work was partially funded by the IOTA Foundation and Science Foundation Ireland.}}

\begin{document}

\title{An adversarially robust data-market for spatial, crowd-sourced data\\

}

\maketitle
\pagestyle{plain}

\begin{abstract}
    We describe an architecture for a decentralised data market for applications in which agents are incentivised to collaborate to crowd-source their data. 
    The architecture is designed to reward data that furthers the market's collective goal, and distributes reward fairly to all those that contribute with their data. 
    We show that the architecture is resilient to sybil attacks, wormhole attacks, and data poisoning.
    In order to evaluate the resilience of the architecture, we characterise its breakdown points for various adversarial threat models in an automotive use case.
\end{abstract}
\section{Introduction}
\label{sec: intro}
\begin{IEEEkeywords}
Smart Cities, Security and Privacy, Service-Oriented Architecture, Crowd Sensing and Crowd Sourcing, Cyber-Physical Systems, Data Management and Analytics, Data markets for mobility applications.
\end{IEEEkeywords}

In recent years there has been a shift in many industries towards data-driven business models \cite{stahl2014data}. Traditionally, users have made collected data available to large platform providers, in exchange for services (for example, web browsing). However, the fairness and even ethics of these business models continue to be questioned, with more stakeholders arguing that such platforms should recompense citizens in a more direct manner \cite{stucke2017should} \cite{8935575}, \cite{fbnytimes}, \cite{aperjis2012market}. This poses many challenges that need to be solved to address new ownership models.

The first one regards fair recompense to the data harvester by data-driven businesses. While it is true that users receive value from companies in the form of the services their platforms provide (e.g., Google Maps), it is not clear that the exchange of value is fair. The second one arises from the potential for unethical behaviours that are inherent to the currently prevailing business models.  Scenarios in which such behaviour have emerged arising out of poor data-ownership models are well documented. Examples of these include Google Project Nightingale, \footnote{\url{https://www.bbc.co.uk/news/technology-50388464}}
\footnote{\url{https://www.theguardian.com/technology/2019/nov/12/google-medical-data-project-nightingale-secret-transfer-us-health-information}} where sensitive medical data was collected of patients that could not opt out of having their data stored in Google Cloud servers. Finally, another challenge is related to the reliability of the data.
Often, the data are generated by users and it is not clear how to protect the system against malicious users that might try to exploit it (e.g., by providing fake data).

The challenges above call for a novel infrastructures to track and trade data ownership. 
These infrastructures can be classified as being either centralised, in which an oracle looks after security and management issues, or decentralised in which trust and security is encoded as part of the data-market peer-to-peer (P2P) protocol. The design of such markets is not new and there have been numerous attempts to design marketplaces to enable the exchange of data for money \cite{stahl2014data2}. This, however, is an extremely challenging endeavour. Data cannot be treated like a conventional commodity due to certain properties it possesses. It is easily replicable; its value is time-dependant and intrinsically combinatorial; and dependent on who has access to the data set. It is also difficult for companies to know the value of the data set a priori, and verifying its authenticity is challenging \cite{agarwal2019marketplace}. Our particular interest is in developing a data-market design that is hybrid in nature; {\em hybrid} in the sense that some non-critical components of the market are provided by trusted infrastructure, but where the essential components of the market place, governing ownership, trust, data veracity, etc., are all designed in a decentralised manner. A detailed description of this hybrid data-market can be found in \cite{kharman2022design}. This data-market is designed for crowd-sourced sensing applications with a view to: (i) achieving a verifiable exchange of data ownership between sellers and buyers; (ii) given an oversupply of data, ensuring that participating agents receive a fair amount of writing access rights to the market; (iii) automatically selecting data points, from all those available, to add the most value to the data-market's collective goal; and (iv)  being robust in adversarial environments by providing protection against Sybil attacks, Wormhole attacks, and Data Poisoning attacks (defined below). Further, we characterise its breakdown points for various adversarial threat models in an automotive use case. Finally, to illustrate the potential utility of the data market, we consider a specific use case in a smart mobility environment that, we hope, might represent a first step towards more general architectures. In this setting, drivers of vehicles within a smart city wish to monetise the data harvested from their cars' sensors \cite{mckinzie2016}.
This use case, while simplifying several aspects, still captures many pertinent aspects of more general data-market design: for example, detection of fake data; certification of data-quality; resistance to adversarial attacks.

To summarize, the contributions of this paper are:
\begin{itemize}
    \item We propose the architecture for a data-market. The architecture is resilient against a number of attack vector and has guarantees of fairness and privacy.
    \item A novel proof-of-work mechanism that is adaptive, useful for the functioning of the market and fair by design. 
    \item We propose an alternative formulation for the Maximum Entropy Voting (MEV) algorithm, called Combination-MEV (C-MEV) and we propose a novel application of this algorithm in the context of this work.  
\end{itemize}

The remainder of this paper is organised as follows: after an overview of related work in Section \ref{RelatedWork}, we present a high-level description of the architecture for the data market in Section \ref{sec: architecture}, as well as describing how each functional component contributes to achieving the desired properties. Then in Section IV we proceed to formalise the definitions used in each component of the data market and describe  each component of the data market in detail. Finally, in section \ref{sec: attacks}, we describe the attacks considered and the resilience of the architecture to those.

\section{Related Work on data-market design} \label{RelatedWork}

Numerous authors have proposed centralized  \cite{travizano2018wibson,agarwal2019marketplace,li2021capitalize,mehta2021sell,rasouli2021data,min2022learn,raja2022market,dawande2022robin} and 
decentralized \cite{ramachandran2018towards,hynes2018demonstration,travizano2018wibson} data markets, recently.
Industry-specific data markets, e.g., in power systems \cite{goncalves2020towards,ACHARYA2022100098}, are also proliferating. A number of recent proposals for data markets use Blockchain architectures as a basis for their work. However, these proposals often fail to address Blockchain design flaws that rapidly become exposed in the context of data-markets \cite{ramachandran2018towards,hynes2018demonstration,travizano2018wibson}. For example, Proof-of-Work (PoW) based Blockchains reward miners with the most computational power. Aside from the widely discussed issue of energy wastage, 
Blockchain based systems also typically use commission based rewards to guide the interaction between users of the network, and Blockchain miners. Such a miner-user interaction mechanism is not suitable in the context of data-markets, effectively prioritising wealthier users' access to the data-market. In addition, miners with greater computational power are more likely to earn the right to append a block, and thus earn the commission. This reward can then be invested in more computational power, leading to a positive feedback loop where more powerful miners become more and more likely to write blocks and earn more commissions. Similarly, the wealthier agents are the ones more likely to receive service for transactions of higher monetary value. This could cause traditional PoW-based Blockchains to centralise over time \cite{beikverdi2015trend}. It is worth noting that centralised solutions to data markets already exist, such as \cite{snowflake}, which namely focus on implementing methods to share and copy data, and to protect certain rights to it, such as read rights. 
Another possible categorisation of prior work relates to the trust assumptions made in the system design. The work in \cite{rasouli2021data} assumes that upon being shared, the data is reported truthfully and fully. In practice, this assumption rarely holds. This assumption is justified in their work by relying on a third party auditor, which the authors of \cite{travizano2018wibson} also utilise. However, introducing an auditor simply shifts the trust assumption to their honest behaviour and forgoes decentralisation.
In \cite{agarwal2019marketplace}, it is identified that the buyer may not be honest in their valuation of data. They propose an algorithmic solution that prices data by observing the gain in prediction accuracy that it yields to the buyer. However, this comes at the cost of privacy for the buyer: they must reveal their predictive task. In practice, many companies would not reveal this Intellectual Property, especially when it is the core of their business model. The work of \cite{narula2018zkledger} is an example of a publicly verifiable decentralised market. Their system allows for its users to audit transactions without compromising privacy. Unfortunately, their ledger is designed for the transaction of a finite asset: creating or destroying the asset will fail to pass the auditing checks. For the transaction of money this is appropriate: it should not be possible to create or destroy wealth in the ledger (aside from public issuance and withdrawal transactions). However, for data this does not hold. Users should be able to honestly create assets by acquiring and declaring new data sets they wish to sell.  Furthermore, their cryptographic scheme is built to transfer ownership of a single value through Pedersen commitments. 
\newline

As we have mentioned, the data-market architecture considered in this paper deviates from those traditionally considered in the literature (such as those mentioned above) in a number of key aspects. First, rather than being centralised, or decentralised, our architecture is hybrid in nature. Secondly, in our-data market, the trust assumptions are embedded in consensus mechanisms that can be verified by all agents using the market place. In other words, the users of a data market have a means to agree on what they trust, and verify that this agreement was reached in a correct and honest manner.\newline

\section{A hybrid data-market architecture}\label{sec: architecture}
Here we present a high-level overview of the proposed architecture. We begin by reminding the reader that the data-market is designed to operate in crowd-sourced environments, where the following conditions prevail:\newline

\begin{figure*}
    \centering
    \includegraphics[width=0.95\textwidth]{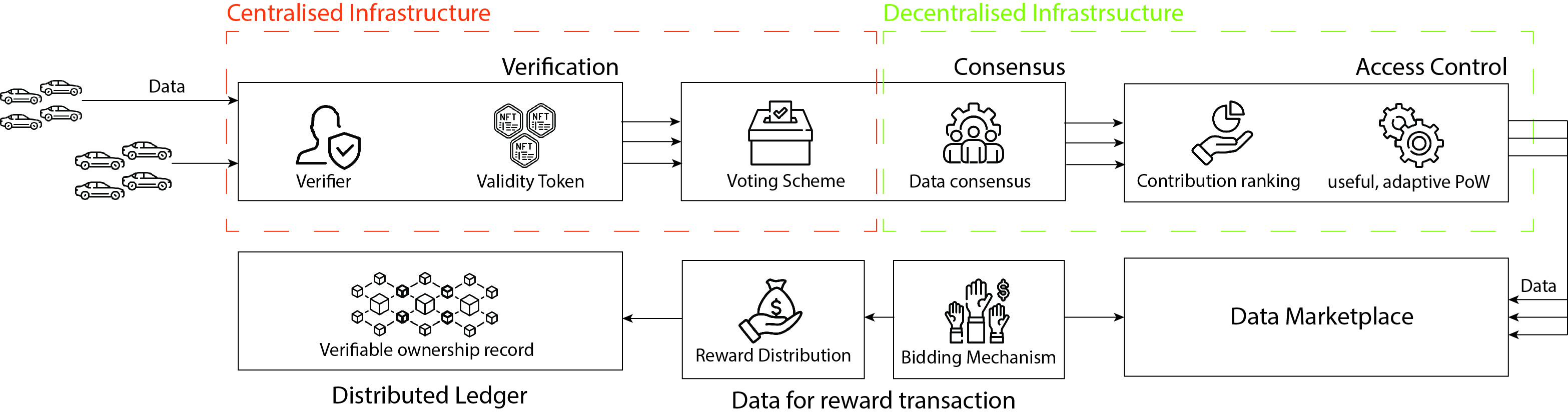}
    \caption{Data Market Architecture. Credit for the images is given in \protect \footnotemark}
    
    \label{fig:Data Market Architecture}
\end{figure*}

\begin{enumerate}
    \item For each potential data point that can made available in the data-market there is an over-supply of measurements, as is the case, for example, in many crowd-sourcing applications. \label{Assumption1}\newline
    \item Competing sellers are interested in aggregating (crowd-sourcing) data points from the market to fulfil a specific purpose. For example, in applications in which sensors measure environmental quantities (rainfall, pollution levels), data may be aggregated to increase robustness, or simply to provide a macroscopic view of the measurements.\newline  
    \label{Assumption2}
    \item Buyers agree to only purchase data from the market, and that each data point in the market has a unique identifier so that replicated data made available on secondary markets can be easily detected by data purchasers. \label{Assumption3}\newline
    \item There is an mechanism that can verify the geographical location of an agent at time of data collection with a certain degree of confidence. 
    One such technique is given in \cite{kharman2023robust}. 
\end{enumerate}
In many mobility applications, where cars are used as sensing devices, many, if not all, of the above conditions prevail. To illustrate this, consider applications where cars seek to monetise information associated with their geographical location. First, agents present a valid proof of their identity and location, as well as demonstrating that their information is timely and relevant. Agents that succeed receive a validity token that allows them to form spatial coalitions with other agents in their proximity. 
Each agent measures data-points from a location quadrant. These data-points are then aggregated by an elected committee of agents from a spatial coalition of that location quadrant. The spatial coalition provides an objective function, that is used to determine the utility of said data.
It is then possible to calculate how valuable data-points are with respect to said objective function. This is done by computing the Shapley value \cite{shapley53} of the data-points. The Shapley value is used to measure the marginal contribution of data-point towards maximising the objective function, and is shown to satisfy notions of fairness.
The higher the Shapley value, the more valuable that data-point is. 
Each agent receives a Shapley value for their data-point provided. This determines the amount of proof-of-work they must compute to sell their data. The greater the Shapley value, the less work they must perform. Indeed, this work consists of computing the Shapley value of a new set of incoming data-points. This feature ensures that spam attacks are costly because for every new data-point an agent wishes to sell, they must perform a new a proof-of-work. Furthermore, the work agents perform is useful for the functioning of the data-market.

An architecture that implements such a system is depicted graphically in Figure \ref{fig:Data Market Architecture}. We now briefly describe the functional components of the data-market.

\paragraph{Verification} 
Agents' position are verified by a proof-of-position mechanism \cite{kharman2023robust} that ensures they provide a valid position and identity. This component ensures that spam attacks are expensive, as well as enabling verifiable centralisation. All agents in the market can verify the validity of a proof-of-position and valid identity because this information is publicly available.

\paragraph{Voting Scheme}
Agents belonging to a spatial coalition agree on what data is worthy of being trusted and sold on the data market. Agents express their preferences for whom they trust to compute the most accepted value of a data point in a given location. This is carried out through a voting scheme.  

\paragraph{Data Consensus}
Once a committee of agents is elected (in the step above), it must come to a consensus as to what is the agreed upon dataset associated to a given location. This is computed by the group following an algorithm that aggregates the coalition's data. This component ensures that at a specific location the dataset does not contain data-points from faulty sensors or from malicious agents (i.e., agents can decide to eliminate outliers).

\paragraph{Access Control Mechanism}
This mechanism used to decide how data should be prioritised to enter the data-market. This mechanism has two steps: firstly, all data-points are assigned a value, which determines the priority they receive to enter the data-market; and secondly, proportionally to this priority, the agent owning that data-point must perform an adaptive, useful proof-of-work to sell their data. 

\footnotetext{In order of appearance: Icons made by Freepik, Pixel perfect, juicy\_fish, srip, Talha Dogar and Triangle Squad from www.flaticon.com}

\paragraph{Data-Market}
The content of the datasets is not public. Buyers purchase the right to access the dataset and perform analytics on them. To provide sufficient information for the buyers, each dataset has metadata associated with it that provides a description of the dataset. The sellers also provide the provenance of the data along with a valid proof-of-position to verify this, and the objective function that their dataset maximises and its value. Buyers can access and browse the market and place bids for specific datasets in exchange for monetary compensation. Buyers may wish to purchase: (i) access (but not ownership) to the entire dataset; (ii) access (but not ownership) to only part of the dataset; (iii) ownership of the dataset or of part of it, which would give them  the rights to redistribute or perform further analytics on said dataset. Each right has a corresponding price that would be decided between the sellers and the buyers.

\paragraph{Distributed Ledger}
Successful transactions are recorded on a distributed ledger to provide a decentralised, immutable record of ownership. The ledger determines which agents have access to particular data, and what access rights they are allowed. Access and ownership to data is managed and exchanged through the use of Non-Fungible-Tokens (NFTs) \cite{battah2022blockchain}.

\section{Components of the Data Market}

 We proceed by describing the algorithmic components proposed in \cite{kharman2022design} and how they are built. Namely, the proposed data market is orchestrated using three key algorithms: one for agent verification; one to implement a reputation-based voting system and an access control algorithm. In this section we describe their technical operation; before proceeding, to aid exposition, we introduce necessary definitions.

\begin{description}
    \item[Data-point]\label{data-point}
    A data-point is defined as $x_i \in X$ where $x_i$ denotes the data point of agent $i$ and $X$ is the space of all possible measurements.
    \item[Location quadrant]\label{location quadrant}
    The set of all possible agent locations is defined as $\mathcal{L} = \{1, 2, 3,...,l \}$. The location quadrant $q$, is an element of  $\mathcal{L}$, where $q \in \mathcal{L}$.
    \item[Buyer]\label{buyer} A buyer is denoted $b$, where $b \in B$ and $B$ is the set of agents looking to purchase ownership (or any other rights) to the datasets available for sale. 
    \item[Spatial Coalition]\label{Spatial Coalition}
    A spatial coalition, $C_q$, is a group of agents in the same location quadrant, $q$.
    \item[Crowdsourced Dataset] Agents in a spatial coalition, $C_q$, aggregate their data-points to provide an agreed upon data-point, $X_t$, at time, $t$.
    \item[Agent]\label{Agent} An agent is defined as $a_i \in A$ where $A$ is the set of all agents competing to complete the marketplace algorithm to become sellers. The index $i \in N$, where $N = |A| $ and $N$ is the total number of agents in the data-market at a given time, $t \in T$.
    \item[Objective Function] An objective function, $v_{C_q}(\cdot)$, maps the aggregate data-points of a spatial coalition to a utility.
    \item [Shapley Value] The Shapley Value, $\psi(C_q)$, is a mechanism to distribute reward amongst a coalition, $C_q$, by considering how valuable the contributions of each agent in the coalition are, with respect to an objective function, $v_{C_q}(\cdot)$. The Shapley Value is the unique reward allocation that satisfies all the properties of the Shapley fairness criteria \cite{shapley53}.

\end{description}
\subsection{The Verification Algorithm}

This algorithm is run by a central authority to provide agents with a seal of credibility for the data they wish to provide. Agents can verify that a commitment \footnote{A cryptographic commitment is a primitive that allows an agent to commit to a value such that it is hidden and cannot be changed later \cite{damgaard1998commitment}.} is well formed and that the proof-of-position algorithm outcome is correct. However, they must trust the central authority to check that the ID provided by the agent is valid \footnote{In future work decentralised proofs of identity may be further explored as a replacement.}. For example in the context of mobility applications, vehicle license plates can be used as a proxy for a valid identity. 

The validity of the data submission must be verified before the data reaches the data marketplace, to avoid retroactive correction of poor quality data. This is done through the $\mathrm{Verifying Algorithm}$\ref{Verfifying Algo}. The following definitions are necessary to understand the functioning of the algorithm:

\begin{description}
    \item[Commitment] \label{Commitment defn}
An agent commits to their data-point by generating a commitment that is hiding and binding, such that the data-point cannot be changed once the commitment is provided. A commitment to a data-point $x_i$, location quadrant $q$, and ID $i$, of an agent $a_i$, at time $t$, is defined as
$\mathnormal{c} \gets \mathrm{Commitment}(a_i, x_i, q, t) $

\item[Proof of ID] \label{PoID defn}
Let the Proof of ID be an algorithm, $\mathrm{ID Proof}$, that verifies the valid identity of an agent $a_i$, with ID $i$. In the context of application that we shall present, this identification will be the license plate of a vehicle. The algorithm will return a boolean, $\alpha$, that will be $True$ if the agent has presented a valid license plate and $False$ otherwise. Then $\mathrm{ID Proof}$ is defined as the following: $\alpha$ $\gets$ $\mathrm{ID Proof}(i)$. It is executed by a central authority that can verify the validity of an agent's identity.

\item[Proof-of-position] \label{PoP defn}
Let proof-of-position be an algorithm, $\mathrm{PoP}$, that is called by an agent $a_i$, with ID $i$. The algorithm takes as inputs the agent's commitment, $c$, and their location quadrant, $q$. We define $\mathrm{PoP}$ as the following algorithm:\\
$\beta$ $\gets$ $\mathrm{PoP}(q, c)$\\
where the output will be a boolean $\beta$ that will be $\mathrm{True}$ if the position $q$ matches the agent's true location and $\mathrm{False}$ otherwise.

\end{description}
\LinesNumbered
\begin{algorithm}[h]
\label{Verfifying Algo}

\SetKw{Return}{return}

$\mathnormal{c} \gets \mathrm{Commitment}(a_i, x_i, q, t) $\;
$\alpha$ $\gets$ $\mathrm{ID Proof}(i)$\;
$\beta$ $\gets$ $\mathrm{PoP}(q, c)$\;
\If{$t$ is not $expired$}{$\gamma$ $\gets$ $True$\;}  
\Else   
    {$\gamma$ $\gets$ $False$\;}

\If{$(\alpha = True, \beta = True, \gamma = True)$}{\Return $True$;}  
\Else   
    {\Return $False$;}

\BlankLine

\BlankLine

\caption{Verification: $\mathrm{Verifying Algorithm}$($a_i$, $x_i$, $q$, $t$, $r$)}
\end{algorithm}

\subsection{Voting Scheme: Reputation-based Maximum Entropy Voting}\label{sec: voting scheme}
As we have mentioned, agents form spatial coalitions to present 
reliable data to the marketplace. We use an adaptation of the Maximum Entropy Voting (MEV) scheme presented in \cite{max_entropy_mackay} that takes into consideration the reputation of agents in the system. This voting scheme reduces the likelihood of selecting an extreme candidate (ie: agent) due to its probabilistic nature. In the context of our work, the selected agents compute the accepted dataset by aggregating data-points.

\subsubsection{Reputation} 

Reputation can be viewed as a trustworthiness metric that is assigned to an agent. Formally an agent $a_i$ assigns a score of trustworthiness to an agent $a_j$. This score is denoted as $r_{i\rightarrow j}$. We assume agents assign reputation following a rational strategy or an agreed upon utility function \footnote{A possible way to allocate reputation could be trusting data measurements based on the age of the vehicle or manufacturer. Other examples include: \cite{ayaz2020voting} and \cite{mahmoud2023framework}.}. 

In the case of Maximum Entropy Voting, which involves solving an optimization problem, the agent running the election must prove that the voting outcome was correctly computed. To provide guarantees of correctness to the voters, we propose using an end-to-end (E2E) verifiable voting scheme. E2E voting schemes require that all voters can verify the following three properties: their vote was \emph{cast as intended}, \emph{recorded as cast} and \emph{tallied as cast} \label{E2Eproperties} \cite{ali2016overview}. To ensure fair and free elections, we require the voting mechanism implementation to also satisfy ballot secrecy as defined in \cite{2015-ballot-secrecy} and \cite{aida2021}. 

\subsubsection{Reputation-based Maximum Entropy Voting}


\begin{defn}[Vote] \label{def:S(a_i)}
The vote of agent $a_i\in A$, is defined as a pairwise preference matrix in  $S(a_i) \in \mathbb{R}^{N\times N}$ . Each entry is indexed by any two agents in $A$ and its value can be derived from data-point $x_i$ and reputation $r_{i\rightarrow j}$. An example of a pairwise preference matrix for three agents is shown in equation \eqref{equ:S(a_i)-example}.
\end{defn}

\begin{defn}[Aggregation of Votes] \label{def:S(A)}
The aggregation of all agents’ votes, $S(A)$, is defined as the average of $S(a_i)$, where $i\in A$, as follows:
\begin{equation}
S(A):=\frac{1}{N}\sum_{a_i\in A} S(a_i).
\label{equ:S(A)}
\end{equation}
\end{defn}

\begin{defn}[Agent Ordering] \label{def:ordering}
An agent ordering, $o$, is defined as a permutation of agents in \cite{max_entropy_mackay}. To reduce computational complexity, we suggest computing $o$ by selecting a subset of agents as the preferred group, such that the order of preferred and non-preferred agents does not matter.
\end{defn}

\begin{defn}[Ordering Set] \label{def:ordering-set}
The ordering set $\mathcal{O}$ is the set of all possible agent orderings, such that $o \in \mathcal{O}$.
\end{defn}


\begin{defn}[Representative Probability] \label{RP}
The \emph{Representative Probability} property states that the probability of the election outcome resulting in candidate A being placed above candidate B should be the same as the proportion of the voters preferring A to B.
\end{defn}

\begin{defn}[Probability Measure of Ordering Set] \label{def:pi}
The (discrete) probability measure, $\pi:\mathcal{O}\rightarrow\mathbb{R}_{\geq 0}$ gives a probability of each ordering $o\in\mathcal{O}$ being selected as the outcome ordering, $o^*$. The probability measure $\pi$, is the one with maximal entropy whilst also adhering to the \emph{Representative Probability} property. $\pi^*$ is the optimal solution to the optimisation problem formulated in equations \ref{pro:MEV0}.
\end{defn}



To combine maximum entropy voting and reputation, a key step is to move from reputation $r_{i\rightarrow j}$ to a pairwise preference matrix $S(a_i)\in\mathbb{R}^{N\times N}$.

\begin{equation}
S(a_1)=
\begin{tabular}{c|ccc}
&$a_i$&$a_j$&$a_k$\\
\hline
$a_i$&0  &1  &1  \\
$a_j$&0  &0  &1/2\\
$a_k$&0  &1/2&0\\
\end{tabular}
\label{equ:S(a_i)-example}
\end{equation}

The entry of a pairwise preference matrix is indexed by every two agents of $A$, and its values are defined as
\begin{equation}
S(a_i)_{j,k}= 
\begin{cases}
1 & \text{if } a_i\text { prefers } a_{j} \text{ and } j \neq k\\
0.5 & \text {if } a_i\text { prefers both equally and } j \neq k\\
0 & \text {if } a_i\text { prefers } a_{k} \text{ or } j=k
\end{cases}
,\label{equ:preference-matrix-agent}
\end{equation}
for $a_j,a_k\in A$. Agent $a_j$ is preferred to $a_k$ if, for example, $\frac{1+\lvert x_i  \rvert \cdot r_{i\rightarrow j}} {1+\lvert x_i - x_j \rvert}
 > \frac{1+\lvert x_i  \rvert \cdot r_{i\rightarrow k}} {1+\lvert x_i - x_k \rvert}
$ and both agents are equally preferred if the two values are equal. 
In this manner, a pairwise preference matrix $S(a_i)$ can be computed for each agent $a_i$. The average of pairwise preference matrices over all agents is denoted as the preference matrix $S(A)$, as described in equation \ref{equ:S(A)}. $S(A)$ represents the pairwise preference of all agents in $A$, whose entries $S(A)_{j,k}$, display the proportion of agents that prefer agent $a_j$ over agent $a_k$.

The original MEV \cite{max_entropy_mackay} runs an optimisation over all candidate orderings, which strongly defines the computational complexity of the problem because the number of orderings is the factorial of the number of candidates.
As a variant of MEV, we consider agent combinations, instead of permutations for the ordering set $\mathcal{O}$, such that $A$ is divided into a preferred group $\mathcal{P}$, of cardinality $K$, and non-preferred group $\mathcal{NP}$. Let $K$ be the number of winners needed and $N = \lvert A \rvert$.  
Hence, the cardinality of the ordering set decreases from $N!$ to $\frac{N!}{K!(N-K)!}$. For small $K$, this leads to to a drastic reduction of the computational complexity. We call this variant of the original MEV, C-MEV (Combination MEV).

For each ordering $o\in\mathcal{O}$, we can define a pairwise preference matrix $S(o)$ (in the same way as \ref{equ:preference-matrix-agent}), whose entries are defined as:
\begin{equation}
S(o)_{j,k}= 
\begin{cases}
1 & \text{if }a_j \text{ is placed over } a_k \\
0.5 & \text {if both are in the same group and } j \neq k\\
0 & \text{if }a_k \text{ is placed over } a_j \text{ or } j=k
\end{cases}
\label{equ:preference-matrix-ordering}
\end{equation}
for $a_j,a_k\in A$.
Let us define an \textbf{unknown} probability measure $\pi: \mathcal{O}\rightarrow\mathbb{R}_{\geq 0}$. $\pi(o),o\in\mathcal{O}$, which gives the probability of $o$ being chosen as the outcome ordering. Then, we construct a theoretical preference matrix $S(\pi)$ as follows:
\begin{equation}
S(\pi):=\sum_{o\in\mathcal{O}} \pi(o)\cdot S(o).
\label{equ:S(pi)}
\end{equation}
The entry $S(\pi)_{j,k}$ states the probability of the outcome ordering placing $a_j$ over $a_k$ under probability measure $\pi$. The definition of \emph{Representative Probability} simply requests that $S(\pi)=S(A)$. 

The entropy of $\pi$ measures the uncertainty of choosing elements in $\mathcal{O}$. The uniform distribution has the maximum amount of entropy. 
Associated with $\pi$, the entropy is defined as $-\sum_{o\in\mathcal{O}} \pi(o) \log \pi(o)$ \cite{gray2011entropy}. Hence, the original formulation of MEV is outlined in \eqref{pro:MEV0}. In this formulation, when maximising entropy, we ensure the solution $\pi^*$ to be the most moderate probability measure which satisfies the \emph{Representative Probability} property.

\begin{equation}
\begin{split}
\pi^*=\max_{\pi} & -\sum_{o\in\mathcal{O}}  \pi(o) \log\pi(o) \\
\text{s.t.} &\sum_{o\in\mathcal{O}}   \pi(o)\cdot S(o) = S(A) \\
&\sum_{o\in\mathcal{O}} \pi(o) = 1 \\
&\pi(o)\geq 0 \quad \forall {o\in\mathcal{O}} 
\end{split}
\label{pro:MEV0}
\end{equation}

$\pi^*$ returns the probability of selecting each ordering $o \in\mathcal{O}$. The result of the C-MEV election computed by randomly sampling an ordering, $o^*$, from the probability distribution $\pi^*$. 

For example: let us consider an election where one winner is selected out of   candidates $A=\{a_1,a_2,a_3\}$. If the probability distribution $\pi^*$, computed following \ref{pro:MEV0}, states that: $\pi^*(o_1) = 0.5$,  $\pi^*(o_2) = 0.3 $  and  $\pi^*(o_3) = 0.2$, where each ordering is defined in Table \ref{tab:ordering-set}, then the election will return agent $a_1$ as the winning candidate with the highest probability. In other words: C-MEV will return the most moderate candidate, whilst ensuring that the outcome adheres to the \emph{Representative Probability} property. 
\begin{table}
\centering
\begin{tabular}{c|ccc}
$\mathcal{O}$&$o_1$&$o_2$&$o_3$\\
\hline
Preferred $\mathcal{P}$&$(a_i)$&$(a_j)$&$(a_k)$\\
Non-Preferred $\mathcal{NP}$&$(a_j,a_k)$&$(a_i,a_k)$&$(a_i,a_j)$\\
\hline
\end{tabular}
\newline
\caption{The lower-carnality ordering set when $A=\{a_i,a_j,a_k\}$ and $M=1$. Agents in the same brackets are given the same rank in an ordering.}
\label{tab:ordering-set}
\end{table}

Finally, since $K$ needs to be kept small to maintain the computational complexity of C-MEV low, in order to increase the amount of candidates, we can perform C-MEV iteratively, $J$ times. This can be done as follows:

\begin{itemize}
    \item Solve C-MEV and find the preferred group $\mathcal{P} = \mathcal{P}_1$, of cardinality $K$.
    \item Obtain $A_1 = A \setminus \mathcal{P}_1$. 
    \item From $A_1$ find the new preferred group $\mathcal{P}_2$, of cardinality $K$.
    \item Repeat the above three steps $J$ times.
\end{itemize}
\begin{figure*}
    \centering
    \includegraphics[width=0.95\textwidth]{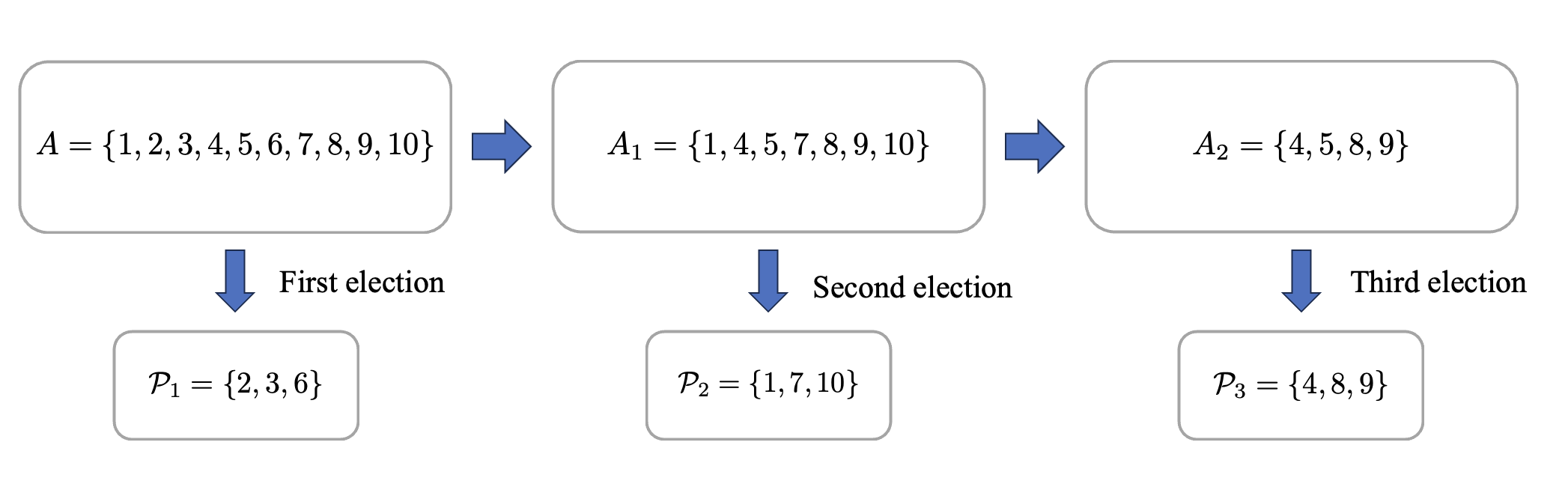}
    \caption{Example of an election performed with C-MEV with parameters $J=3$, $K = 3$.}
    
    \label{fig: Election Example}
\end{figure*}

Refer to Figure \ref{fig: Election Example} for a simple example of this process, with $J=3$ and $K = 3$. 

This iterative version of C-MEV allows us to obtain $J$ groups of preferred candidates of dimension $K$. Notice that this is a drastic decrease in computational complexity with respect to solving C-MEV for groups of $K\cdot J$ candidates. In fact, the computational complexity to solve C-MEV for a group of $K\cdot J$ candidates is proportional to $\frac{N!}{(K\cdot J)!(N-K\cdot J)!}$, whereas the computational complexity to solve C-MEV, $J$ times for groups of $K$ candidates is 
\begin{eqnarray}
    \sum_{i = 0}^J \frac{(N - K\cdot i)!}{K!(N- K\cdot(i+1))!} < J\frac{N!}{K!(N-K)!} < \nonumber\\< \frac{N!}{(K\cdot J)!(N-K\cdot J)!} \nonumber
\end{eqnarray}

\subsection{Data Consensus}\label{sec: data consensus}

Coalitions of agents in the same location quadrant must agree on the accepted value of measured data. The rationale for this algorithmic component is that, since we are assuming to be in environments with an oversupply of data, we can use the large amount of measurements to filter out data-points that have been submitted either by malicious agents or agents with faulty sensors. The underlying assumption is that the majority of the agents participating in the data collection are honest (i.e., they will not report fraudulent or fake data). It is also assumed that agents in the same geographic area measure similar results within margins of measurement precision. Therefore, by aggregating agents' measurements, the aim  of the Data Consensus component is to resolve conflicts in diverging results. Moreover, the aforementioned aggregation should be computed in a privacy-preserving manner. This prevents agents from stealing valuable data-points from other agents.

To recap, the Data Consensus component needs to satisfy three properties: (i) it needs to be executed in a decentralised fashion; (ii) untrue data-points submitted by malicious actors or faulty sensors should not appear in the final dataset of the spatial coalition; (iii) it must computed be in a privacy-preserving manner, such that malicious agents are prevented from stealing valuable data-points from other agents.

We now introduce two concepts that we use to characterise different algorithms and measure how they achieve the three aforementioned properties.

\begin{itemize}
    \item \emph{K-Privacy}: a decentralised algorithm in which $N$ agents provide individual inputs $x_i$ to compute an output $\Tilde{x}$ is $k$-private if it is not possible to reconstruct the dataset $X = \{x_1, x_2, ...\}$ unless $k$, or more agents collude with one another by exchanging information, (ie: their $x_i$'s) where $k$ is a positive integer. In other words, if a dataset satisfies k-privacy with $k = 3$, at least three agents must share their $x_i$'s to reconstruct $X$.
    
    \item \emph{Breakdown Point:} In estimation theory, the breakdown point characterises the robustness of an estimator and is dependent on the sample size $n$ \cite{rousseeuw1985multivariate}. 
    In our context the \emph{theoretical breakdown point} of the Data Consensus algorithm is the minimum share of agents required to alter the dataset of the spatial coalition arbitrarily \cite{breakdown_point}. The \emph{practical breakdown point} is the average share of agents required to alter the dataset of the spatial coalition arbitrarily.
\end{itemize}

With these definitions we may characterise the algorithm of choice for the Data Consensus mechanism: a decentralised mean-median algorithm. The rationale for this choice is that the mean-median algorithm represents a compromise between a decentralised computation of Mean and a decentralised computation of median. The former can be computed in a decentralised way \cite{Overko2019} and is n-private, where $n$ is the number of agents in the spatial coalition. However its theoretical breakdown point is $\frac{1}{n}$. In other words, a single agent is required to arbitrarily alter the dataset of the spatial coalition. The latter, on the other hand, is one of the most robust estimators as its theoretical breakdown point is $\frac{1}{2}$. However, computation of the median in a privacy-preserving way is a computationally complex task \cite{median_privat}.

The mean-median algorithm is an algorithm that can estimate the median value of the dataset in a robust and private manner.

The first step of the algorithm is to randomly assign every agent to a group in such a way that there are $g$ groups with at least $s$ agents each. $g$ and $s$ determine the properties of privacy and robustness of the algorithm. The next step is to calculate the mean within each group. The resulting mean is at least of k-privacy: $k = s-1$. The value of each mean is then broadcast among each group and it is then possible to compute the median of all the means. As there are $g$ groups, there are $g$ ways in which the median can be chosen. As such, the theoretical breakdown point, $P$, of the mean-median algorithm is:
\begin{equation}\label{eq: breakdown point mean-median}
   P = \frac{g}{2n}
\end{equation}
The relationship between $s$, $g$ and the number of agents $n$, is given by the following inequality:
\begin{equation}\label{eq: mean-median}
    n \geq s \cdot g \newline
\end{equation}









\subsection{Access Control Mechanism to the Data Market}

A spatial coalition, $C_q$, is formed by agents within the same location quadrant, where $C_q = \{a_{q_1}, a_{q_2}, a_{q_3}...\}  \subseteq A$ and $ \{q_1, q_2, q_3...\}$ are the IDs of the agents in quadrant $q$. Each agent $a_{q_i} \in C_q$ measures a data-point $x_{q_i}$ at a given time $t$. Agents in $C_q$ agree on an objective function, $v_{C_q}(\cdot)$ which determines the utility of their dataset. A committee of agents is elected from agents in $C_q$ using Maximum Entropy Voting, following section \ref{sec: voting scheme}. This committee aggregates data-points provided by agents in $C_q$ using an aggregation algorithm outlined in \ref{sec: data consensus}. The resulting aggregated data-point is denoted as  $X_t$, where $X_t = \emph{mean-median}(x_{q_1}, x_{q_2}, x_{q_1}...)$. 
It is then possible to calculate how valuable data-points are with respect to said objective function. This is done by computing the Shapley value, $\psi(X_t)$. The Shapley value is used to measure the marginal contribution of data-point towards maximising the objective function, and is shown to satisfy notions of fairness. Agents in $C_q$ must complete a proof-of-work that is inversely proportional to the Shapley value they received for their data-point contribution. The more valuable their data, the less work they must perform \footnote{In this context, more work would involve computing the Shapley value of a greater number of data-points.}. Indeed, this assigned proof-of-work consists of computing the Shapley Value of the next incoming data-point(s) wishing to enter the data-market, $\psi(X_{t+1})$. This work enables the functioning of the market. More specifically, we make use of smart contracts \footnote{A smart contract is a program that will automatically execute a protocol once certain conditions are met. It does not require intermediaries and allows for the automation of certain tasks \cite{7467408}\cite{szabo1994}.} to automate this process. In our context, a smart contract will be executed by one agent $a_{i}$ to compute the Shapley value of agent $a_{j}$’s data. The outputs will be the Shapley value of agent $a_{j}$’s data and a new smart contract for agent $a_{i}$. Calculating the new smart contract generated serves as the proof of agent $a_{j}$’s useful work.

\begin{figure*}
    
    \includegraphics[width=\textwidth]{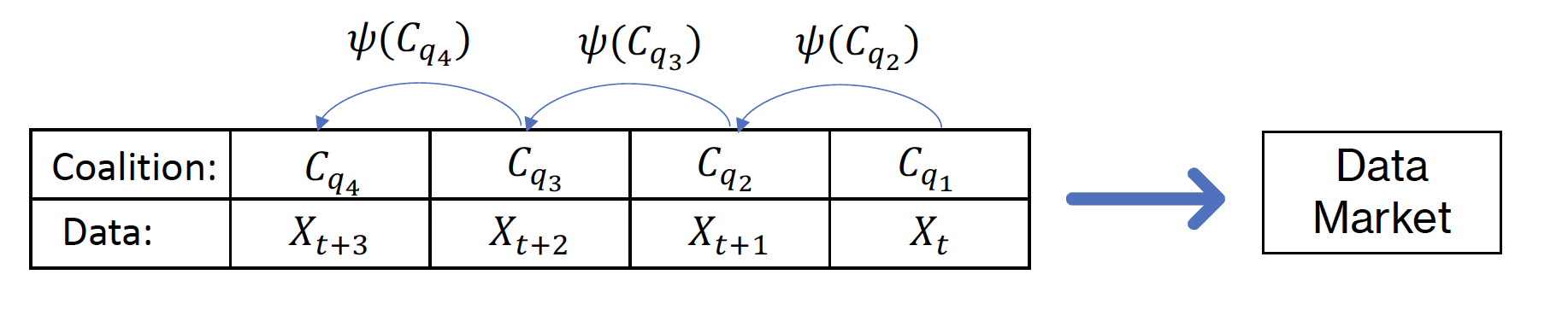}
    \caption{Access Control Mechanism: Agents in coalition $C_{q_1}$ must compute the Shapley value, $\psi(C_{q_2})$, of the new incoming data, $X_{t+1}$, using $X_{t+1}$ and $v_{C_{q_1}}(\cdot)$. Once the agents of $C_{q_2}$ receive their Shapley value, they must complete an amount of proof-of-work that is inversely proportional to their Shapley value. This work is to compute the Shapley value of $\psi(X_{t+2})$ and so on. Once a coalition completes this work they may enter the market.}
    
    \label{fig:Access Control Mechanism}
\end{figure*}

\subsubsection{A contextual example: Crowd-sourcing pollution measurements}

We illustrate an example with real data to showcase how the Shapley value would be used to rank the data-points in terms of value, and allocate a proportional proof-of-work correspondingly.

We use the data on pollution levels of a range of different contaminants, taken from a number of cities in India. The data has been made publicly available by the Central Pollution Control Board \footnote{\url{https://cpcb.nic.in/}}. The cleaned and processed data was accessed from \footnote{\url{https://www.kaggle.com/datasets/rohanrao/air-quality-data-in-india}}. We illustrate an example wherein a buyer is interested in purchasing data on pollution levels of different contaminants in order to predict the Air Quality Index (AQI) of a given location. We generate a linear regression model to predict AQI, which has been previously done in \cite{linearregAQI}, although other options for models to predict AQI have been explored in alternative works such as \cite{ameer2019comparative} and \cite{kumar2011forecasting}. This model will serve as the objective function.

A description of how AQI is calculated can be found in \footnote{\url{https://app.cpcbccr.com/ccr_docs/How_AQI_Calculated.pdf}}. Following from this calculation, it is reasonable to observe how the variables PM2.5 (Particulate Matter 2.5-micrometer in $\mathnormal{\mu g / m^3}$) and PM10 (Particulate Matter 10-micrometer in $\mathnormal{\mu g / m^3}$) are highly correlated with AQI. We include them as well as NO, NO2, NOx, NH3, CO, SO2 and O3 as training features for the linear regression model.

Agents collecting measurements of different pollutants have their data-points evaluated by a preceding set of agents that must calculate some proof-of-work. This proof-of-work is computing the Shapley value of a data-point. To do so, they utilise the given objective function, which in this case is the linear regression model, and the agent's data.

\begin{figure*}[h]
    \centering
    \includegraphics[width=0.75\textwidth]{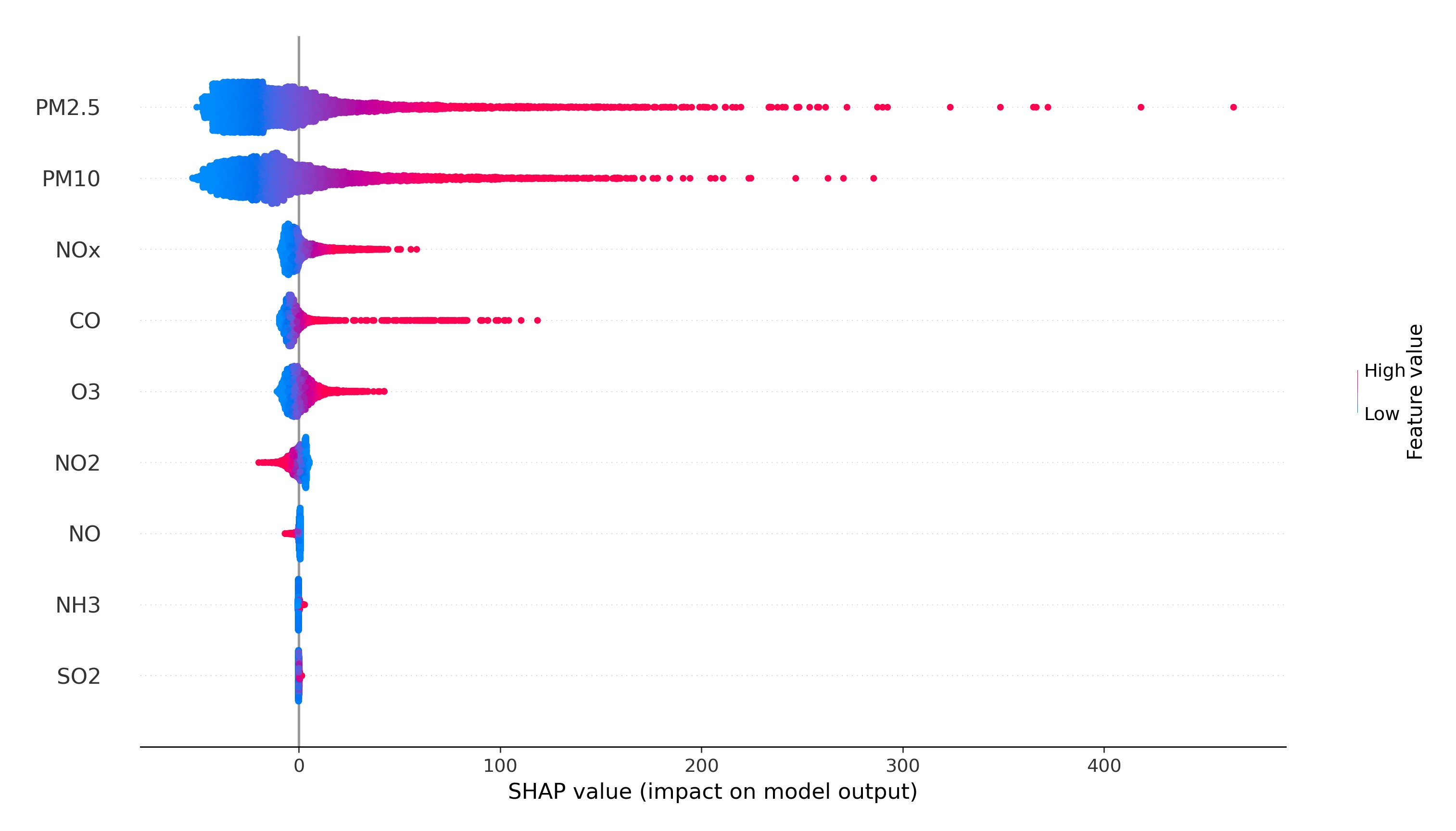}
    \caption{SHAP Values of the samples of each feature}
    \label{fig:scatterplot}
\end{figure*}

We show the results of calculating the Shapley value of individual data-points within a given dataset in Figure \ref{fig:scatterplot}. We simulate this using the SHAP library, presented in \cite{NIPS2017_7062}. Following from the SHAP documentation: "Features pushing the prediction higher are shown in red, those pushing the prediction lower are in blue" \cite{shapdocs}. By visual inspection of Figure \ref{fig:scatterplot}, we can conclude that data-points measuring high PM2.5 and PM10 concentrations increase the predicted AQI the most. 

Consequently, the agents having provided those data-points would receive the highest Shapley values, and thus have to perform less proof-of-work. In this context, that would mean computing the Shapley value of a smaller number of new incoming data-points.



\begin{remark}
    The reader may rightly question the privacy risks of an agent accessing another one's dataset to compute the Shapley value. What is to incentivise them to compute the Shapley value honestly, and what is to prevent them from stealing or duplicating another agent's data if they realise it has a high Shapley value? 

    To address the first concern, the computation of the Shapley value is automated through the use of smart contracts. These can be inspected by anyone to ensure their correct operation.

    Secondly, in the market there is no protection against agents duplicating data, but they cannot monetise this copied data unless they go through the verification, consensus and then access control stages again. Because we are in an environment with an oversupply of data and that is crowd-sourced, the data is unlikely to be highly sensitive and thus the incentive to go through these steps is very small. In addition to this, if data duplication is indeed deemed a pertinent issue for the context of a different data market implementation, the Shapley Value calculation can be replaced by the Shapley Robust algorithm proposed in \cite{agarwal2019marketplace}, for which we have written a Python implementation using the SHAP explainer Python library. This is an adaptation of the Shapley Value calculation that will penalise highly similar data, thereby penalising attackers wishing to monetise duplicated data. This implementation can be found in \url{https://github.com/aidamanzano/DataMarket}.
\end{remark}


\subsection{Purchasing Datasets}
Once the steps above have been completed and the metadata of data have been posted publicly on the market, the sellers chose which buyer or buyers they will sell their dataset to. Finally, once a transaction is successful, the reward of the sale is distributed amongst the participants of the spatial coalition that generated the dataset according to a given reward distribution function. 

\section{Robustness of data-market in adversarial environments} \label{sec: attacks}
In an adversarial, decentralised environment, one must take into account the possibility of attacks on the system. We proceed to describe their nature and how these are mitigated by the algorithmic components of the data market architecture.

\begin{defn}[Sybil Attack]
Sybil Attacks are a type of attack in which an attacker creates a large number of pseudonymous identities which they use to exert power in and influence over the network.
\end{defn}
Sybil attacks are mitigated in the verification stage, as agents must present a valid proof of identity. This proof is granted to them through a centralised authority but all other agents can verify that it exists and therefore that it must be valid.
In the context of smart mobility applications, generating multiple identities is made expensive because agents must provide a valid vehicle license plate to enter the market and sell data. 

\begin{defn}[Wormhole Attack]
A Wormhole Attack involves a user maliciously reporting they are in a location that is not the one they are truly in.
\end{defn}
An attack can be mounted by a series of malicious actors claiming to measure data from a location they are not truly in, and wishing to monetise this fraudulent data. To mitigate against this attack, agents must present a valid proof-of-position in the verification stage. This proof is assumed to be correct and sound, and by definition, agents are only able to present one valid proof.

\begin{defn}[Data Poisoning]
Data Poisoning is an attack where malicious agents collude to report fake data in order to influence the agreed upon state of a system.
\end{defn}
Malicious agents wishing to report fake data must influence enough agents in their spacial coalition to ensure that sufficient agents in the Data Consensus mechanism get elected to compute a fake data-point. Probabilistic voting schemes make the cost of this coercion significantly high. Furthermore, to sell the uploaded data-point, the agent must first perform a useful proof-of-work that is proportional to how valuable the data point is deemed. The less useful the data point the more work the agent must complete to sell it. Selling spam data will therefore be very time consuming for an attacker.


\begin{figure*}
    \centering
    \includegraphics[width=0.8\textwidth]{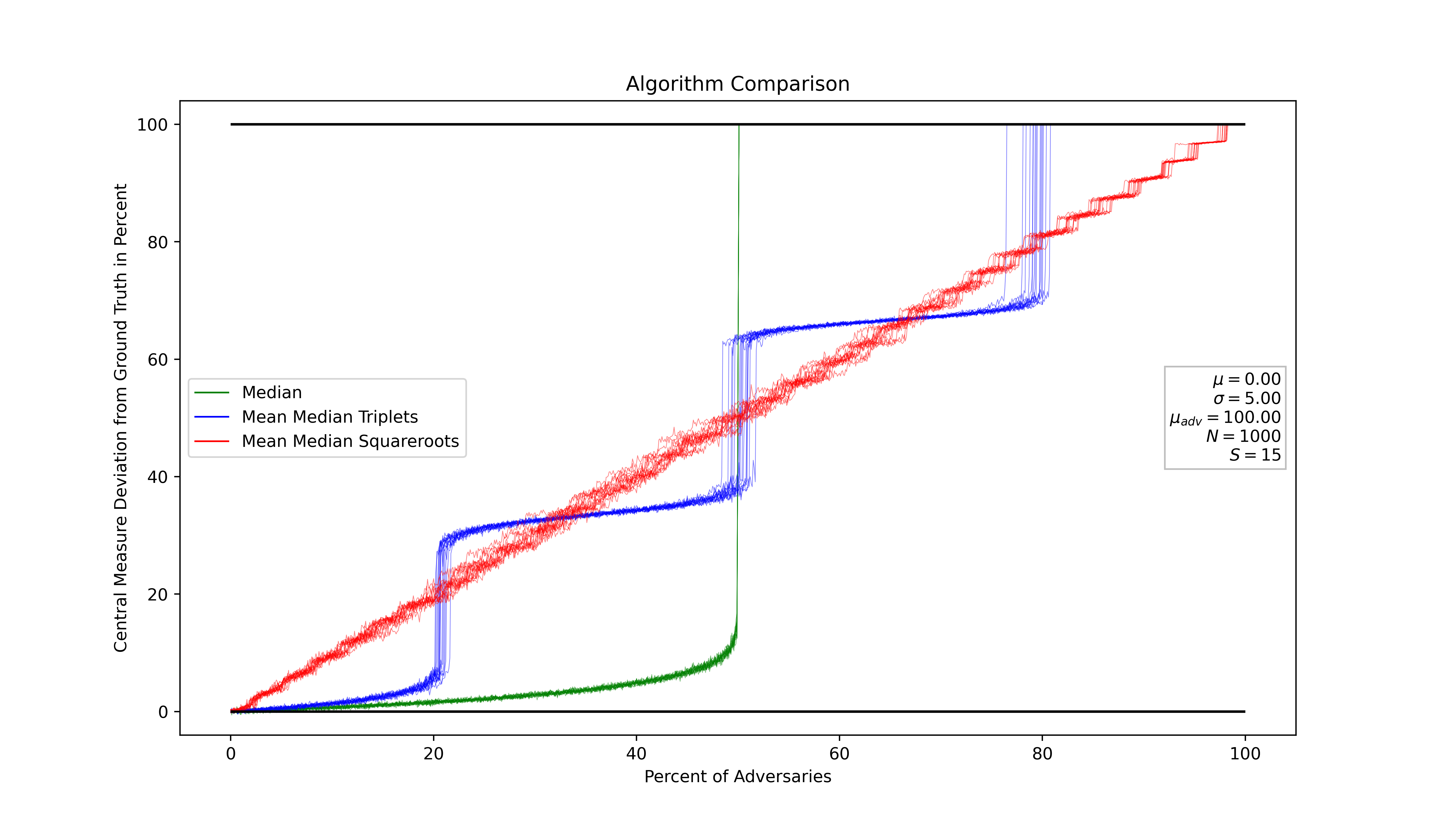}
    \caption{Characterisation of data consensus algorithms' behaviour under different degrees of coordinated data poisoning attacks.}
    \label{fig:algo characterisation}
\end{figure*}

\begin{figure*}
    \centering
    \includegraphics[width=0.8\textwidth]{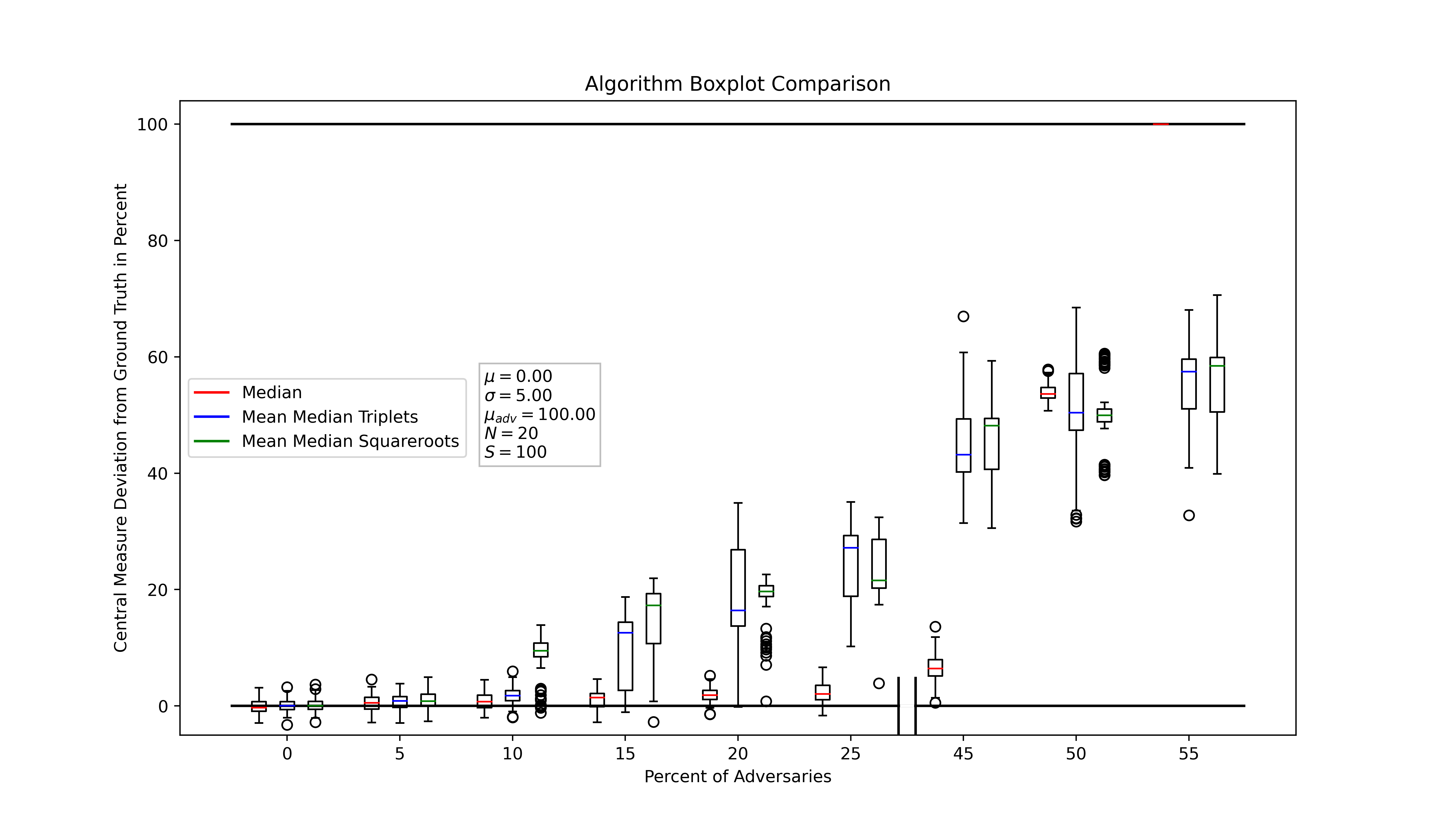}
    \caption{Breakdown analysis of data consensus algorithms, with a coordinated data poisoning attack}
    \label{fig:algoboxplot}
\end{figure*}

As the \emph{Sybil attack} and the \emph{Wormhole attack} are not possible due to the design of the data-market, we proceed by showcasing results that illustrate the robustness of the data-market with respect to the \emph{Data Poisoning} attack.



\subsection{Simulation Setup}

We consider a number of agents $N$ taking measurements from the same location quadrant. To account for faulty sensors and other sources of errors, the process of taking a measurement is represented by sampling from a Gaussian distribution with mean, $\mu$, and standard deviation, $\sigma$. Here we assume that the value $\mu$ represents the \emph{true} value of the phenomenon that agents are measuring. Furthermore, we assume that a set of collaborative malicious agents report a fake measurement $\mu_{adv}$. This is done to simulate a \emph{Data Poisoning} attack.

Each agent is randomly assigned a reputation score: with probability $\frac{1}{2}$ they are assigned a value of $1$, otherwise the reputation is sampled from a Gaussian distribution with $\mu_{rep}$ and $\sigma_{rep}$, where $\mu_{rep}$ is a value much larger than $1$. 

We consider two scenarios to test the mean-median algorithm: firstly with group size, $s$, of agents  $s=3$ (triplet group size), and secondly with sizes chosen depending on the number of agents, $N$, with $s=\sqrt{N}$. 

For each scenario we perform $S$ number of Monte-Carlo simulations, in which we vary the numbers of agents, $N$, and the size of the malicious coalition. Furthermore, we show results not only for the mean-median algorithm but also for the decentralised mean and the decentralised median, to showcase the difference in performance.


\subsection{Evaluation of Results}\label{sec: simulation}
Figure \ref{fig:algo characterisation} shows a simulation of the mean-median algorithm for the Data Consensus mechanism described in Section \ref{sec: data consensus}. We also show the results for the median algorithm to provide a baseline for comparison. This simulation has been computed with $S=25$ and $N=1000$. Notice that the number of agents $N$ is greatly exaggerated with respect to a realistic scenario and its results are functional to establish a baseline behaviour in the presence of a large number of agents. In green we show the behaviour of the decentralised median algorithm, in blue the behaviour of the mean-median algorithm with $s = 3$ and in red the behaviour of the mean-median algorithm with $s = \sqrt{N}$, as the size of the adversarial coalition increases.

It can be observed that, when the percent of adversaries in the network increases, the deviation of the reported data-point from the ground truth increases. For the median algorithm, this deviation is significant when the percent of adversaries in the network is 50\%, which is consistent with the theoretical breakdown point. For the mean-median algorithm with $s=3$ (triplets), there are three different percentages at which a significant deviation from the ground truth can be observed. This is because when the percentage of adversaries is below 20\%, it is likely that most agents within the triplet groups will be honest. For a percent of adversaries between 20\% and 50\%, it is most likely that at least one out of three is malicious. For a percentage between 50\% and 80\%, two out of three are likely to be malicious and finally, when the percentage of adversaries is above 80\%, it is most probable that all agents in the median group are malicious. 

Regarding the mean-median algorithm with $s = \sqrt{N}$, (square-roots), it shows a continuous, almost linear, relation to the number of adversaries present. This is a result of a larger group size, with $s=\sqrt{N}$ and $N=1000$, which allows for a more uniform distribution of honest and malicious agents in the median groups.

\begin{figure*}
    \centering
    \includegraphics[width=0.75\textwidth]{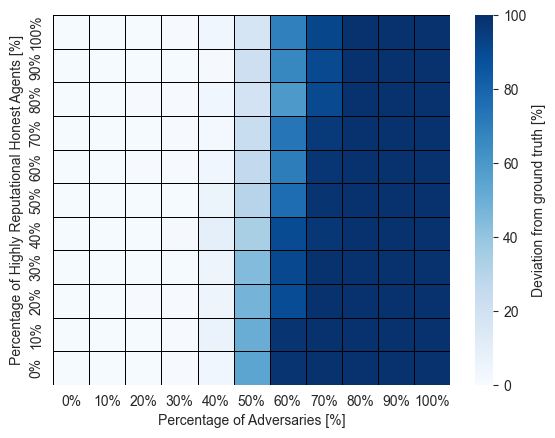}
    \caption{Characterisation of breakdown of C-MEV combined with mean-median algorithm}
    \label{fig:MEV medianboxplot}
\end{figure*}



To investigate the scaling effect and behaviour of the algorithm when smaller numbers of agents are present (which also represents more realistic scenarios), the same simulation was carried out but with $N=20$ agents, as depicted in Figure \ref{fig:algoboxplot}. The number of Monte-Carlo simulations is increased to $S=100$ to provide meaningful statistical results. 

From the results shown in Figure \ref{fig:algoboxplot}, it can be observed that the algorithms act similarly. Given a greater percent of adversaries present in the system, the deviation of the reported data from the ground truth increases. This is a direct result of the lower number of agents, where chances of random fluctuations are higher. When the number of agents decreases, the practical breakdown point increases from 3\% to 10\% for the square-roots algorithm, and for the triplets algorithm it decreases from 20\% to 15\%. These results are in accordance with the theoretical breakdown point defined in \eqref{eq: breakdown point mean-median}. 

Finally, Figure \ref{fig:MEV medianboxplot} represents how much the reported data deviates from the ground truth, when combining the reputation-based C-MEV voting scheme and the Data Consensus algorithm. The results obtained demonstrate that the combination of both mechanisms provides an increased robustness against \emph{Data Poisoning} attacks, assuming a functional reputation system exists. In this simulation the voting scheme (C-MEV) outputs $J = 5$ sets of agents of cardinality $K=3$, out of a set of $N=30$ agents, which then aggregate data-points to provide the agreed upon data-point of a given location, according to the mean-median algorithm. 

Figure \ref{fig:MEV medianboxplot} is a heatmap that depicts the results of the simulations. The y-axis represents the percentage of highly reputational agents within the set of honest agents, and the x-axis represents the percentage of adversaries. The number of Monte-Carlo simulations is set to $S=100$ to provide meaningful statistical results. We sample the reputation for the high-reputation honest actors, from a gaussian distribution with parameters $\mu_{rep}=100$ and $\sigma_{rep}=30$.

Visual inspection of the proposed results lead to two main observations: (i) combining C-MEV and data consensus has a higher practical breakdown point than data consensus alone; (ii) an increased share of highly reputational agents among the honest ones, leads to an increase of the practical breakdown point. This is because, the higher the reputation of a honest agent, the more likely they are to be selected, thus decreasing the chances of malicious agents being elected. 

To conclude, it can be said that combining Maximum Entropy Voting and the \color{black} mean-median\color{black}, the system offers strong protection against \emph{Data Poisoning}. To succeed in mounting a \emph{Data Poisoning} attack, the malicious coalition must control from 40\% to 60\% of the network, assuming a functional reputation system exists.

\section{Conclusions}

We have presented a fair, decentralised data-market architecturethat is robust to a number of attacks. The novelty of this work includes: ranking data in terms of how valuable it is using the Shapley value, and using it to proportionally adapt the proof-of-work each agent must complete. Furthermore, the proof-of-work is itself useful and necessary for the functioning of the market, and thus not wasteful. We also utilise a voting scheme that satisfies desirable properties of fairness, and introduce an optimisation to make its computational complexity significantly lower for the context of this work. 
We evaluate the resilience of the Data Consensus algorithm combined with the voting mechanism towards \emph{Data Poisoning} attacks. Our simulations show an increased robustness. 

\textbf{Acknowledgements:} The authors would like to thank the IOTA Foundation for funding this research and Juan Antonio Vera García for the idea of allowing sellers to provide the objective function.

{
\footnotesize \bibliographystyle{unsrt}
\bibliography{bibliography}
}

\clearpage
\onecolumn

\end{document}